\documentclass{article}
\usepackage{spconf,amsmath,graphicx}
\usepackage{subfigure}
\usepackage{amsthm,amsmath,amssymb}
\usepackage{cite}
 \usepackage{multirow}
\usepackage{mathrsfs}

\title{Few-shot Medical Image Segmentation via Cross-Reference Transformer}
\name{Yao Huang\textsuperscript{1}, Jianming Liu\textsuperscript{1,}$^{\ast}$ and Hua Chen \thanks{$^{\ast}$Corresponding author: Jianming Liu(liujianming@jxnu.edu.cn). \textsuperscript{1}These authors contributed equally to this work.  This work was financially supported by the Natural Science Foundation of China (No.62266022 and No.61662034), the Natural Science Foundation of Jiangxi Province(20202BAB202020) and the Jiangxi Double Thousand Plan (JXSQ2019101077). } }
\address{Jiangxi Normal University, Nanchang, China}

\begin{document}
%
\maketitle

\begin{abstract}
Deep learning models have become the mainstream method for medical image segmentation, but they require a large manually labeled dataset for training and are difficult to extend to unseen categories. Few-shot segmentation(FSS) has the potential to address these challenges by learning new categories from a small number of labeled samples. The majority of the current methods employ a prototype learning architecture, which involves expanding support prototype vectors and concatenating them with query features to conduct conditional segmentation. However, such framework potentially focuses more on query features while may neglect the correlation between support and query features. In this paper, we propose a novel self-supervised few shot medical image segmentation network with Cross-Reference Transformer, which addresses the lack of interaction between the support image and the query image. We first enhance the correlation features between the support set image and the query image using a bidirectional cross-attention module. Then, we employ a cross-reference mechanism to mine and enhance the similar parts of support features and query features in high-dimensional channels. Experimental results show that the proposed model achieves good results on both CT dataset and MRI dataset.
\end{abstract}
\section{Introduction}
\label{sec:intro}
With the rapid development of AI technology in recent years, medical clinical applications have been more widely used and play a pivotal role in modern medical diagnosis. Automatic segmentation of medical images is the core technology of many medical clinical procedures, which directly affects the accuracy of medical diagnosis. With the development in recent years, medical image segmentation technology based on traditional fully supervised deep learning has been approaching maturity\cite{U-Net,Unet++,2018Attention,nnUnet,TransUNet,UneXt,SwinUnet}, and has achieved excellent results. Although these methods have achieved excellent segmentation results, they are often impractical in practical applications. First, the annotation of medical images is very difficult to obtain. It requires a large number of experts in the field of medical imaging and requires them to spend a lot of time to annotate, which is extremely expensive. Second, there are differences in the acquisition procedures of different medical devices, and there are differences in the images collected by different medical devices, which makes some announcement datasets may be difficult to meet the needs of hospitals or some medical institutions. Third, the generalization ability is too poor. The fully supervised model is inflexible when facing any new segmentation target (anatomical structure or lesion) that requires the collection of annotated data for the new class and retraining of the model. 

In order to solve the problem of limited number of image data and low generalization, few-shot learning technology has been proposed\cite{Prototypical,Learning,Few-Shot,Matching,One-ShotObject,One-shotSimple}. In the forward inference process of few-shot segmentation, the few-shot segmentation model only needs a small amount of support label information to extract the discriminative representation of unseen classes, and then it can predict the segmentation of unseen classes in the query image. There is no need to label the new class of images and retrain the model. As can be seen from the characteristics of small sample learning, it has great potential in medical image segmentation problems.

In recent years, few-shot image segmentation (FSS) has made great progress, and it is even as good as fully supervised image segmentation. However, most few-shot segmentation models can only be applied to the segmentation of natural images\cite{Adaptive, Part-aware,SSP,NERTNet}, because there are modal differences between medical images and natural images, as the contrast of medical images is usually low. The few-shot segmentation model that performs well in natural images cannot be directly used on medical images. Although the few-shot model has strong generalization and can directly predict the segmentation of unseen classes, the few-shot model still requires a large amount of labeled data information in the training stage, and there are still reasons for the difficulty in obtaining annotations for medical images, which leads to a small amount of data. These reasons make the research on few-shot medical image segmentation models develop very slowly in recent years. The existing medical image few-shot segmentation models are basically based on the 2D segmentation framework. SENet\cite{SENet} is the first few-shot model that uses few-shot techniques for medical image segmentation. Ouyang et al proposed self-supervised FSS framework for medical images (SSL-ALPNet)\cite{SSL-ALP}, which uses superpixels for eliminating the need for manual annotations, and exploits an adaptive local prototype pooling enpowered prototypical network (ALPNet). SSL-ALPNet has played a milestone role in promoting fewshot medical image segmentation, because it uses superpixel self-supervised technology and has achieved very excellent results, which makes it no longer difficult to train fewshot medical image segmentation models due to the lack of medical image data. 

Despite the promising performance of above proto -type based methods, there are still several drawbacks. Firstly, the cooccurrent features between support images and query images are critical to making accurate segmentation decisions on query images; however, this is currently being ignored by existing methods. Secondly, current prototypical networks do not sufficiently account for the interaction between support features and query features during the training phase. This lack of interaction can lead to the failure of generating fully representative prototypes.
In this work, we present a novel self-supervised few-shot medical image segmentation network with cross-reference transformer. Our method addresses the lack of interaction between the existing Cross-reference support image and the query image, and can better mine and enhance the similar parts of support features and query features in high-dimensional channels. We adopt same self-supervised learning framework as SSL-ALPNet and use superpixels\cite{Optimal,2003Learning,Efficient} to train few-shot segmentation model, which not only solves the problem of lack of annotation in medical images, but also generalizes well in practice.
Our contributions are as follows:

(1) We propose a novel Cross-Reference Transformer Prototype Network (CRTPNet) that can be used for few-shot medical image segmentation

(2)  A Cross-Reference Transformer block is proposed, which can mine and enhance similar parts between support and query features to enhance the foreground.

(3) Our proposed method achieves advanced performance on CT dataset and MRI dataset.
\section{Related Work}
\subsection{Medical Image Segmentation}
With the development of computing hardware, deep learning methods have entered the field of computer vision and have begun to demonstrate their powerful image processing capabilities. In recent years, significant progress has been made in medical image segmentation using deep neural networks. Based on CNN, networks were trained and tested on human and brain MRI to segment bones and tumors from the background\cite{2017CNN}. Subsequently, the fully convolutional network (FCN) replaced the last fully connected layer in CNN with convolutional layers, enabling the network to perform dense pixel-level prediction. inspired by FCN \cite{2015Fully}, Ronneberger et al. \cite{U-Net} introduced a famous medical image segmentation network called U-Net. It is based on FCN \cite{2015Fully} and uses many feature channels during upsampling to allow the network to pass context information to higher resolution layers. Then, many U-Net related improved methods were proposed. For example, Zhou et al. combined the Densenet-like structure to propose Unet++\cite{Unet++}, which improved the gradient mobility through dense skip connections. Oktay et al. proposed Attention U-Net \cite{2018Attention} to enhance the segmentation accuracy of the model for medical images by adding an attention module to skip connections. Isensee et al. proposed nnU-Net\cite{nnUnet} based on the traditional U-Net model by improving the process of data preprocessing and data augmentation. And with the popularity of Vision Transformer, TransUnet \cite{TransUNet} combining U-Net and Vision Transformer and Swin-Unet \cite{SwinUnet} with a similar structure to U-Net have appeared in the past two years. However, these methods require a large amount of annotated data to fully unleash their potential and lack the ability to perform segmentation prediction on new categories. 
\subsection{Few-shot Segmentation}
Few-shot segmentation (FSS) refers to training a model that can quickly segment new classes by introducing additional support prior knowledge on the basis of few-shot learning. PANet\cite{PANet} is a classical fewshot segmentation model. It uses prototypes to segment the prediction image and proposes prototypes alignment regularization, which makes full use of the knowledge of support. In the last two years, SSP\cite{SSP} proposed a method to extract query prototypes to segment query images. NERTNet\cite{NERTNet} proposes a model to eliminate background and interfering regions in a query image. Although few-shot segmentation is still a hot development direction, it is mostly in the segmentation of natural images, and it develops slowly in the field of medical image segmentation. The main reason is that although the few-shot segmentation model can quickly identify new classes, it requires a large amount of labeled data in the training phase, and medical image annotations are difficult to obtain. However, many researchers have proposed solutions. SENet \cite{SENet} is the first few-shot segmentation model for medical images, which uses a channel squeezing and spatial excitation module to enhance the interaction between the two branches for image segmentation. In order to solve the problem of insufficient data, SSL-ALPNet\cite{SSL-ALP} proposes to use the superpixel self-supervised method to train the model, and has a profound influence on many subsequent studies. Qinji et al. proposed a location-sensitive local prototype network by introducing spatial prior knowledge\cite{2021A}. CRAPNet\cite{CRAPNet} proposes a cyclic similarity attention module to enhance the segmentation of medical images. 
\section{PROPOSED METHOD}
\label{sec:format}

\begin{figure}[htb]
\begin{center}
\includegraphics[width=8.65cm,height=3.3cm]{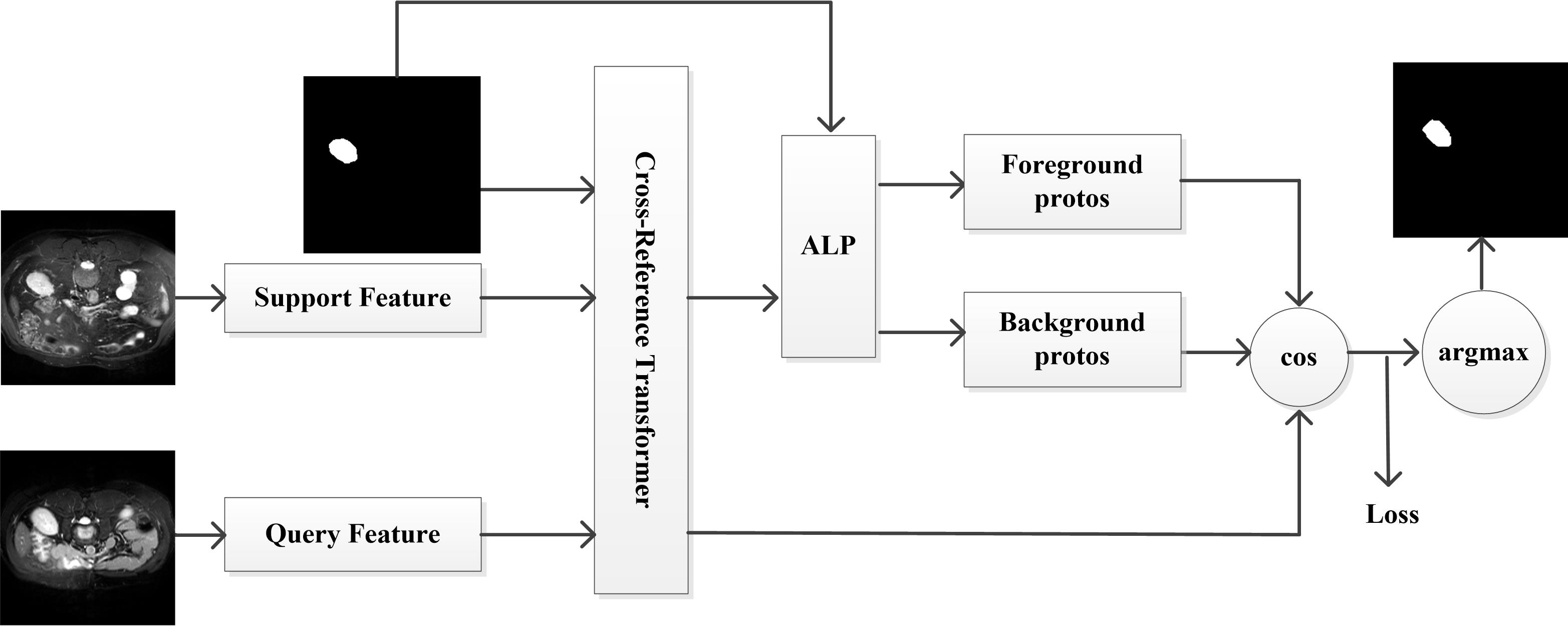}
\caption{Overflow of Proposed Network}
\end{center}
\end{figure}

\begin{figure*}[htb]
\begin{center}

\subfigure{
\includegraphics[width=9.55cm,height=3.3cm]{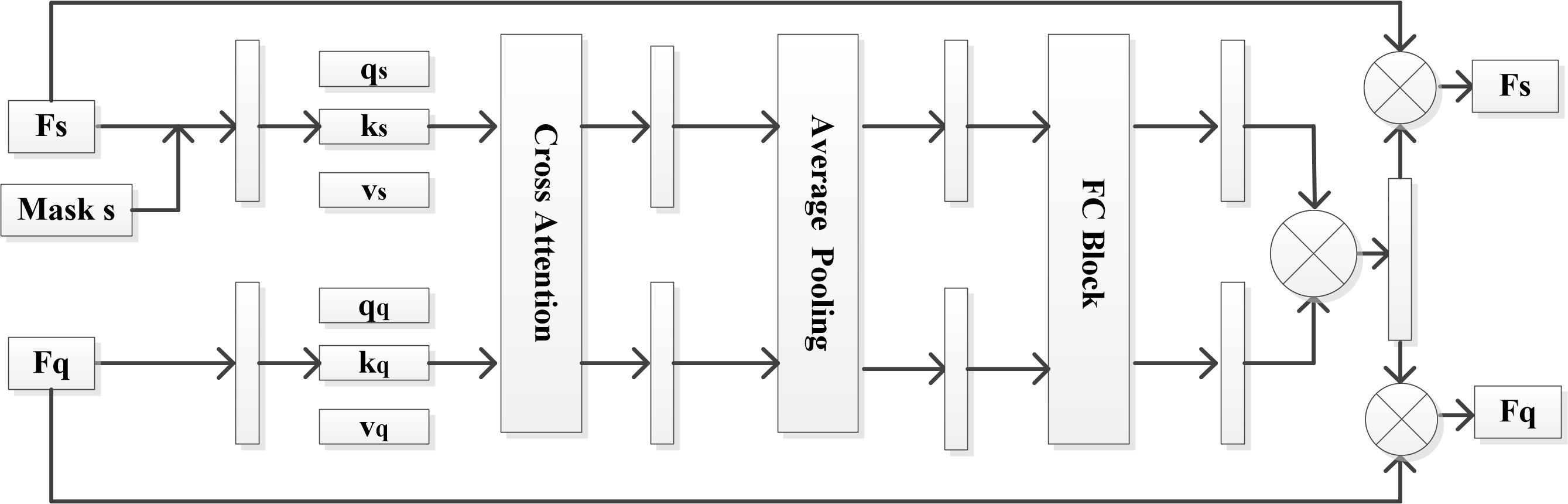}
}
\caption{The proposed Cross-Reference Transformer Block}
\end{center}
\end{figure*}

We first introduce the few-shot segmentation problem, then introduce the proposal process and overall structure of Cross-Reference Transformer, and finally introduce the network structure of CRTPNet.

\subsection{Problem Definition}
Few-shot segmentation enables the model to segment unseen classes with few annotated examples. In the usual setting of few-shot semantic segmentation, the dataset is usually divided into a training set $D_{train}$ containing the training class $C_{train}$ and a test set $D_{test}$ containing the test class $C_{test}$, and $C_{train}$ and $C_{test}$ do not contain a common class, that is, $C_{train} \cap C_{test}=\varnothing$. And the background class is not considered, that is, the background class is not included in $C_{train}$ and $C_{test}$. The episodic training is usually adopt in the few-shot segmentation, where the model is trained with many epochs and one epoch contains many episodes. During training, each $episode$ $i$ contains a support-query pair ($S_{train}$, $Q_{train}$), where each $S_{train}$=($X_{s}$, $Y_{s}$) is the support set and $Q_{train}$=($X_{q}$, $Y_{q}$) is the query set. $X$ refers to the image, and $Y$ denotes the corresponding mask annotation. Through training, the model will eventually acquire the ability to extract new class discriminations from the support set and predict the segmentation of the query image.

\subsection{Network Architecture}
The overall flow of our network is shown in Figure. 1. The framework includes three components: (1) a basic feature extraction module, which is used to extract the feature maps $f_{\theta}(x^{q}) {\in} \mathbb {R}^{D {\times} H \times W}$ and $f_{\theta}(x_{l}^{s}) {\in} \mathbb {R}^{D \times H \times W}$ from the query image and the support image, where  ($H$,$W$) is the spatial size, $D$ is the channel depth, and $l$ is the prototype index; (2) The proposed cross-reference transformer module, which are feed with the features and the mask annotations of the supporting images;(3) a prototype and similarity based classifer for segmentation. We use an Adaptive local prototype pooling(ALP)\cite{SSL-ALP} module to obtain the prototype. And then we perform the similarity strategy to obtain the foreground mask and background mask, and the final segmentation result is obtained after argmax of the results.

\subsection{Cross-Reference Transformer}
Inspired by \cite{2020CRNet}, we propose a Cross-Reference Transformer  block, which is a powerful module that can mine out and enhance the similar parts of supporting features and query features. Its working flow is shown in Figure 2. Where $F_{s}$ represents the feature map of the support image,  $F_{q}$ represents the feature map of the query image, Average Pooling represents the global pooling, FC Block contains two linear layers, a ReLU activation function and a Sigmoid function. Different from natural images, the foreground of medical images only accounts for a small part of the image and the background similarity is very high, so in order to eliminate the interference of background and other features, the support mask is firstly used to multiply with the support image, which can better mine the foreground features in the query image. Then, a bidirectional cross attention is done, where the $q_{s}$ is the query feature from the support image and the $k_{q}$, $v_{q}$ are key and value features from the query image. They are used to calculate the support-to-query  attention feature, the calculation formula is shown as Formula (1), where $d$ is the dimension of the input sequence. And the query feature $q_{q}$ from support image with $k_{s}$, $v_{s}$,  from the query image are used to calculate the query-to-support attention feature, shown as Formula (2). The two intermediate features obtained are then globally pooled to acquire the global statistics in the two images. Then, the two feature vectors are sent to a pair of two-layer fully connected (FC) Block to obtain two attention weights. After that, the two attention weights obtained by $F_{s}$ and $F_{q}$ through FC Block are fused by element-wise multiplication. At this point, the module has mined the high-level similar features related to the foreground existing in $F_{s}$ and $F_{q}$ to obtain the final attention weights. Finally, the obtained attention weight is multiplied by the original respective feature maps, and the reinforced support feature $F_{s}$ and $F_{q}$ query feature  are obtained. Different improvements are shown in the chapter on ablation experiments.

\begin{equation}
    Attention(q_{s}, k_{q}, v_{q}) = softmax(\frac{q_{s}k_{q}^{T}}{\sqrt{d}})v_{q}
\end{equation}

\begin{equation}
    Attention(q_{q}, k_{s}, v_{s}) = softmax(\frac{q_{q}k_{s}^{T}}{\sqrt{d}})v_{s}
\end{equation}

\subsection{Prototype and Similarity-based Segmentation}
We use the ALP method\cite{SSL-ALP} to generate local prototypes and compute class-level prototypes within modules. The ALP module can preserve local information in the prototype, specifically by using the pooling window size ($L_{H}$,$L_{W}$) for every $f_{\theta}(x_l^s) {\in} \mathbb {R}^{D \times H \times W}$. Compute the local prototypes $p_{l, _mn}(c)$ for class $c$ at each average intersection feature map spatial location ($m$, $n$). It is represented by the following equation.

\begin{equation}
    \begin{split}
        p_{l, mn(c)} = avgpool(f_{\theta}(x_l^s))(m, n)\\
        = \frac{1}{L_HL_W}\sum\limits_{h}\sum\limits_{w}f_{\theta}(x_l^s)(h, w)
    \end{split}
\end{equation}
$where\quad mL_H < h < (m + 1)L_H, nL_W < w < (n + 1)L_W$\\

After obtaining the local prototypes, in order to avoid too small models not being acquired, a class-level prototype $p_l^g(c^j)$ is calculated, using average pooling [23], where $y_l^s(c^j)$ is the binary mask, and $j$ represents the category except the background.

\begin{equation}
    p_l^g(c^j) = \frac{\sum\limits_{h,w}y_l^s(c^j)(h,w)f_{\theta}(x_l^s)(h,w)}{\sum\limits_{h,w}y_l^s(c^j)(h,w)}
\end{equation}
Finally, local prototypes and class-level prototypes are grouped together by concatenation operation to form a prototype set $P = \{p(c^j)\}$.

To predict the class $Y_{q}(h,w)$, we compute the cosine similarity and get the prediction using following formulas, where $\alpha$ is a multiplier with a value of  20 to help with backpropagation.

\begin{equation}
    S_{c^j}(h, w) = \alpha p(c^j)\odot f_{\theta}(x_l^q)
\end{equation}

\begin{flalign}
\begin{split}
Y_q(h,w) = softmax(S_{c^j}(h,w) softmax(S_{c^j}(h,w)))
\end{split}
\end{flalign}

\subsection{Training strategy}
In each training iteration, a support-query pair is selected based on the strategy described in previous work\cite{SSL-ALP}. Specifically, the original image is used as the support image, and the query image is obtained by applying affine transformation $\mathcal{T}_g$ and gamma transformation $\mathcal{T}_i$ to the original image. In order to simulate the situation in reality, where the amount of medical image data is lacking, the experiment is set as a 1-way-1-shot setting. And the pseudo-labels generated by superpixels are input into the network as the corresponding label data for training. We use the commonly used cross-entropy loss as the loss function for conducting experiments. And referring to the idea of prototype alignment\cite{PANet}, that is, the prediction result of the query image is used as the support image, and the support image is re-segmented. For each $episode$  $i$ we use the following formula to calculate the loss.

\begin{flalign}
\begin{split}
&\mathcal{L}_{seg}^i(\theta;\mathcal{S}_i,\mathcal{Q}_i)=\\
&-\frac{1}{HW}\sum\limits_{h}^H\sum\limits_w^W\sum\limits_{j\in\{0,p\}}\mathcal{T}_g(y_i(c^j))(h,w)log(\hat{y}_i(c^j)(h,w))
\end{split}
\end{flalign}

\begin{flalign}
    \begin{split}
        &\mathcal{L}_{reg}^i(\theta;\mathcal{\hat{S}}_i,\mathcal{S}_i)=\\
        &-\frac{1}{HW}\sum\limits_{h}^H\sum\limits_w^W\sum\limits_{j\in\{0,p\}}
y_i(c^j)(h,w)log(\widetilde{y}_i(c^j)(h,w))
    \end{split}
\end{flalign}

$y_i(c^j)$ here is a binary mask, $\hat{y}_i(c^j)$ is the prediction of query, $\widetilde{y}_i(c^j)$ is the prediction of taking original support image as query. We add these two loss values to get the final loss value, $\lambda$ is the control strength of regularization and is set to 1. The final loss is as follows:

\begin{equation}
    \mathcal{L}_{seg}^i(\theta;\mathcal{S}_i,\mathcal{Q}_i)=\mathcal{L}_{seg}^i + \lambda\mathcal{L}_{reg}^i
\end{equation}
\section{Experiments}
\label{sec:pagestyle}

\begin{table*}[!ht]

\center
\begin{tabular}{c|c|c|c|c|c|c|c|c|c|c}\hline
\multicolumn{1}{c|}{\multirow{2}{*}{Method}} & \multicolumn{5}{c|}{Abdominal MRI} & \multicolumn{5}{c}{Abdominal CT} \\ \cline{2-6} \cline{7-11}
 & LK  & RK & Spleen & Liver & Mean & LK  & RK & Spleen & Liver & Mean \\ \hline

 SENet\cite{SENet}& 45.78 & 47.96  & 47.30 & 29.02 & 42.51 & 24.42 & 12.51 & 43.66 & 35.42 & 29.00  \\ \hline
 PANet\cite{PANet}& 30.99 & 32.19 & 40.58 & 50.40 & 38.53 & 20.67 & 21.19 & 36.04 & 49.55 & 31.86  \\ \hline
 SSL-PANet\cite{SSL-ALP}&  58.83 &  60.81 &  61.32 &  71.73 &  63.17 & 56.52 & 50.42 & 55.72 &  60.86 &  57.88  \\ \hline
 SSL-ALPNet\cite{SSL-ALP}& 81.92 & 85.18 & 72.18 & 76.10 & 78.84 & 72.36 & 71.81 & 70.96 & 78.29 & 73.35  \\ \hline
 SSL-RPNet\cite{2021Recurrent}& 71.46 & 81.96 & 73.55 & 75.99 & 75.74 & 65.14 & 66.73 & 64.01 & 72.99 & 67.22  \\ \hline
 CRAPANet\cite{CRAPNet}& 81.95 & 86.42 & 74.32 &76.46 & 79.79 &74.69 & 74.18 & 70.37 & 75.41 & 73.66  \\ \hline
 Ours& $\pmb{82.81}$ & $\pmb{87.34}$ & 73.82 & $\pmb{76.75}$ & $\pmb{80.18}$ & 73.63 & 71.27 & $\pmb{73.97}$ & $\pmb{77.30}$ & $\pmb{74.04}$  \\ \hline
\end{tabular}
\caption{Experiment results on Setting 1 (in Dice score (\%))}
\end{table*}

\begin{table*}[!ht]

\center
\begin{tabular}{c|c|c|c|c|c|c|c|c|c|c}\hline
\multicolumn{1}{c|}{\multirow{2}{*}{Method}} & \multicolumn{5}{c|}{Abdominal MRI} & \multicolumn{5}{c}{Abdominal CT} \\ \cline{2-6} \cline{7-11}
 & LK  & RK & Spleen & Liver & Mean & LK  & RK & Spleen & Liver & Mean \\ \hline

 SENet\cite{SENet}& 62.11 & 61.32  & 51.80 & 27.43 & 50.66 & 32.83 & 14.34 & 0.23 & 0.27 & 11.91  \\ \hline
 PANet\cite{PANet}& 53.45 & 38.64 & 50.90 & 42.26 & 46.33 & 32.34 & 17.37 & 29.59 & 38.42 & 29.43  \\ \hline
 SSL-PANet\cite{SSL-ALP}&  47.71 &  47.95 &  58.73 &  64.99 &  54.85 & 37.58 & 34.69 & 43.73 &  61.71 &  44.42  \\ \hline
 SSL-ALPNet\cite{SSL-ALP}& 73.63 & 78.39 & 67.02 & 73.05 & 73.02 & 63.34 & 54.82 & 60.25 & 73.65 & 63.02  \\ \hline
 CRAPANet\cite{CRAPNet}& 74.66 & 82.77 & 70.82 &73.82 & 75.52 &70.91 & 67.33 & 70.17 & 70.45 & 69.72  \\ \hline
 Ours&$\pmb{76.74}$  & 80.15 & 70.07 & 73.36 & 75.08 & 66.37 & 61.05 & 67.92 & $\pmb{73.88}$ & 67.31  \\ \hline
\end{tabular}
\caption{Experiment results on Setting 2 (in Dice score (\%))}
\end{table*}

\begin{figure}[htb]
\begin{center}
\subfigure{
\rotatebox{90}{\scriptsize{~~~~~~~~~~~Support}}
\includegraphics[width=2cm,height=2cm]{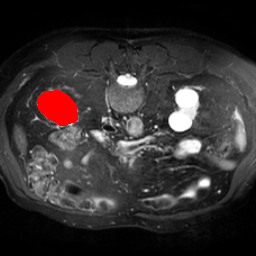}
\includegraphics[width=2cm,height=2cm]{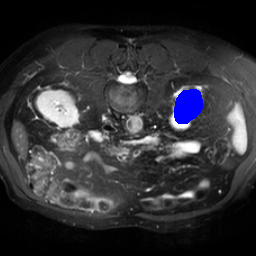}
\includegraphics[width=2cm,height=2cm]{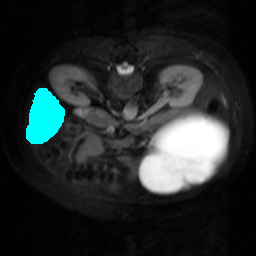}
\includegraphics[width=2cm,height=2cm]{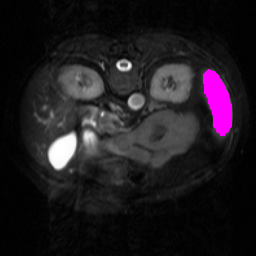}
}
\subfigure{
\rotatebox{90}{\scriptsize{~~~~~~~~~~~Label}}
\includegraphics[width=2cm,height=2cm]{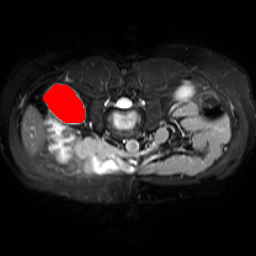}
\includegraphics[width=2cm,height=2cm]{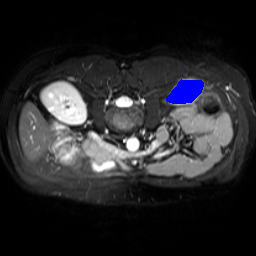}
\includegraphics[width=2cm,height=2cm]{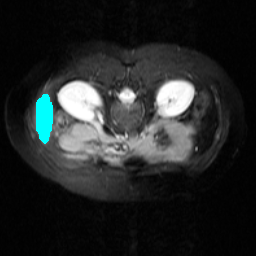}
\includegraphics[width=2cm,height=2cm]{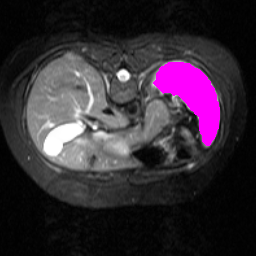}
}
\subfigure{
\rotatebox{90}{\scriptsize{~~~~~~~~~~~Proposed}}
\includegraphics[width=2cm,height=2cm]{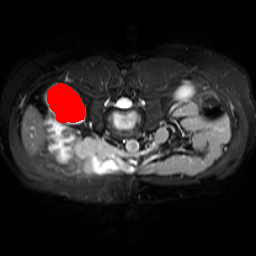}
\includegraphics[width=2cm,height=2cm]{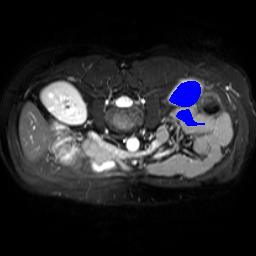}
\includegraphics[width=2cm,height=2cm]{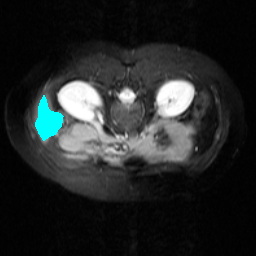}
\includegraphics[width=2cm,height=2cm]{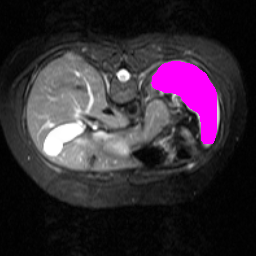}
}
\caption{Example of prediction of CRTPNet on the MRI dataset on Setting 1}
\end{center}
\end{figure}

\begin{figure}[htb]
\begin{center}
\subfigure{
\rotatebox{90}{\scriptsize{~~~~~~~~~~~Support}}
\includegraphics[width=2cm,height=2cm]{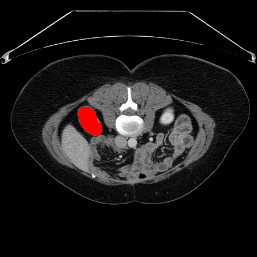}
\includegraphics[width=2cm,height=2cm]{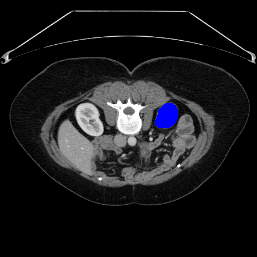}
\includegraphics[width=2cm,height=2cm]{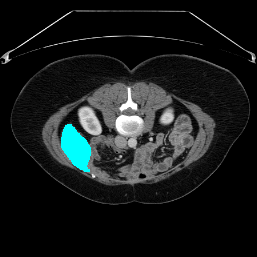}
\includegraphics[width=2cm,height=2cm]{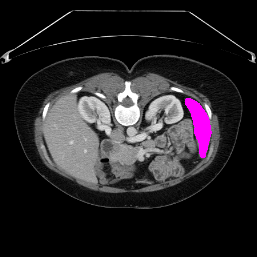}
}
\subfigure{
\rotatebox{90}{\scriptsize{~~~~~~~~~~~Label}}
\includegraphics[width=2cm,height=2cm]{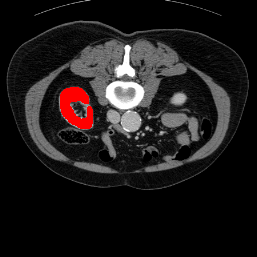}
\includegraphics[width=2cm,height=2cm]{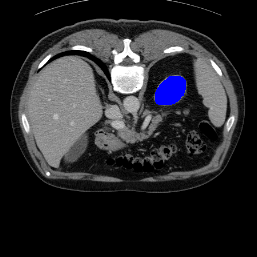}
\includegraphics[width=2cm,height=2cm]{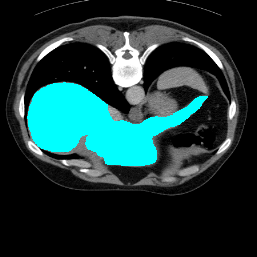}
\includegraphics[width=2cm,height=2cm]{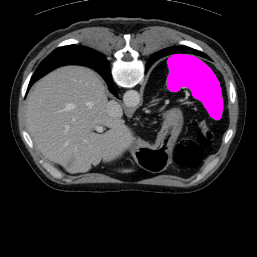}
}
\subfigure{
\rotatebox{90}{\scriptsize{~~~~~~~~~~~Proposed}}
\includegraphics[width=2cm,height=2cm]{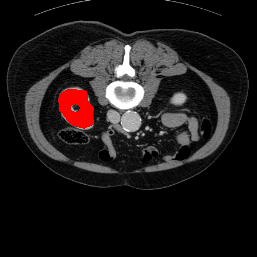}
\includegraphics[width=2cm,height=2cm]{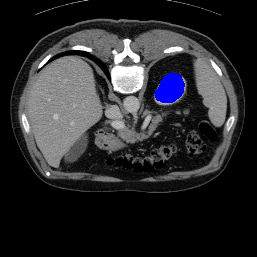}
\includegraphics[width=2cm,height=2cm]{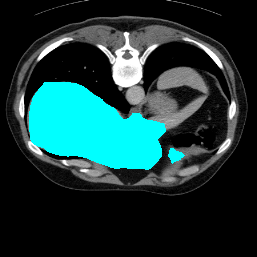}
\includegraphics[width=2cm,height=2cm]{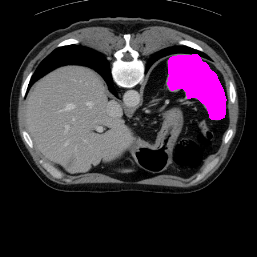}
}
\caption{Example of prediction of CRTPNet on the CT dataset on Setting 1}
\end{center}
\end{figure}
\subsection{Datasets}
To assess the performance of the proposed method, we conducted tests on MRI and CT datasets, followed by experimental evaluations for each dataset. The abdominal MRI dataset used in our study is the CHAOS-T2 dataset from the ISBI2019 Joint Healthy Abdominal Organ Segmentation Challenge (Task 5), comprising 20 3D T2-SPIR MRI scans\cite{2021CHAOS}. The abdominal CT dataset utilized in our research is derived from the MICCAI 2015 Multi-Atlas Abdomen Labeling Challenge, consisting of 30 3D abdominal scans\cite{CT}.

The experiment is conducted using a 2D model framework, where 2D slices from the dataset are utilized for training, with the image size uniformly set to 256 × 256. As per the standard approach, we replicate the image across its channel three times, resulting in an image data size of 3 × 256 × 256, which serves as the input for the model.

Five-fold cross validation is used to evaluate the performance and only 1-way-1 shot setting is used. Liver, spleen, left kidney and right kidney in both datasets are used as semantic class sets. In addition, in order to better compare with other advanced methods, Setting1 and Setting 2 \cite{SENet,SSL-ALP} set by predecessors are used in all experiments. Setting1 is set so that unseen classes can appear in the background class of the training set, Setting2 is set so that unseen classes cannot appear in the background class of the training set, and the dataset is divided into upper abdominal group and lower abdominal group. To demonstrate the effectiveness of the model, we compare with other state-of-art methods.

\subsection{Evaluation metric}

The evaluation index uses DSC coefficients commonly used in medical images. The DSC coefficient is used to calculate the similarity of two sets, that is, to evaluate how similar the predicted results are to the label, and its value ranges from 0 to 1, with larger values indicating that the predicted results are more similar to the label.

\subsection{Implementation Details}

We use the Pytorch library to code the implementation of the proposed model. The backbone extraction network, using ResNet101 pre-trained on MS-COCO dataset for higher segmentation performance\cite{PANet}. Adam was used as the optimizer for training with an initial learning rate of 0.001 for 100,000 iterations, where every 1,000 iterations the learning rate decays to the original 0.98 with a batch size of 1. The experimental environment uses a piece of NVIDIA GeForce RTX3090.

\subsection{Comparison with the state-of-the-art methods}

In order to assess the performance of the proposed method, we compare the performance of the proposed model with excellent models in recent years. Table 1 and Table 2 shows the comparison results of different models in setting1, while Table 3 and Table 4 shows the comparison results of different models in setting2 on the data set in recent years. PANet\cite{PANet} is the most widely influential few-shot model in the field of few-shot image segmentation on natural images. SENet \cite{SENet} is the first few-shot segmentation model proposed for medical images. SSL-ALPNet\cite{SSL-ALP} introduced the milestone of using superpixel self-supervision to train few-shot medical image models. RPNet\cite{2021Recurrent} is a supervised method with a recursive mask optimization module to iteratively optimize the segmentation mask, \cite{CRAPNet} adapt it into the same self-supervision learning framework and applies setting1 to it and denoted as SSL-RPNet; CRAPNet\cite{CRAPNet} is the latest SOTA model for 2023. Compared with CRAPNet, our method outperforms most of the state-of-the-art models and only slightly outperforms CRAPNet. Figures 3 and 4 show examples of the model's segmentation predictions on different datasets, respectively; The first row is the support map, the second row is the label map, and the third row is the segmentation prediction of the model.

\subsection{Ablation experiment}
The Cross-Reference Transformer is inspired by the \cite{2020CRNet}. To explore the promotion made by different improvements, we do ablation experiments as shown in Table 5. Base is Cross-Reference i.e. there is no additional Support Mask input and no Transformer cross-reference module. From the ablation experiment results, we can see that after the additional introduction of the Support Mask, the performance improvement is still very low despite the exclusion of the interference of irrelevant features. This paper believes that this is caused by too little interaction between support features and query features, and then the performance is greatly improved after adding the cross Transformer, which also proves the effectiveness of this work.

\begin{table*}[!ht]

\center
\begin{tabular}{c|c|c|c|c|c}\hline
\multicolumn{1}{c|}{\multirow{2}{*}{Method}} & \multicolumn{5}{c}{Abdominal MRI}  \\ \cline{2-6} 
 & LK  & RK & Spleen & Liver & Mean \\ \hline

 Base & 81.78 & 84.40  & 73.29 & 76.26 & 78.93  \\ \hline
 Base+Support Mask & 80.76 & 85.93 & 73.75 & 76.10 & 79.14  \\ \hline
 Base+Mask+Transformer& 82.81 & 87.34 & 73.82 & 76.75 & 80.18\\ \hline
\end{tabular}
\caption{Ablation Experiment results on Setting 1 (in Dice score (\%))}
\end{table*}
\section{CONCLUSION}
In this paper, we propose novel self-supervised few shot medical image segmentation network with cross-reference transformer, which addresses the lack of interaction between the support image and the query image, and can better mine and enhance the similar parts of support features and query features in high-dimensional channels. Our model achieves state-of-the-art performance on the CHAOS-T2  dataset and  the abdominal CT dataset.





\bibliographystyle{IEEEbib}
\bibliography{refs}

\end{document}